\documentclass[12pt]{article}
\usepackage{graphicx}
\begin{document}
version 22.7.2013
\begin{center}
{\Large Solution of the relativistic bound state problem for hadrons}  
\vspace{0.3cm}

H.P. Morsch\footnote{permanent address: HOFF, Brockm\"ullerstr.~11,  
D-52428 J\"ulich, Germany\\ E-mail:
h.p.morsch@gmx.de }\\
National Centre for Nuclear Research, Pl-00681 Warsaw,
Poland
\end{center}

\begin{abstract}
A second order extension of the QED Lagrangian (including boson-boson coupling)
has been used to describe $q\bar q$ hadrons. Assuming
massless elementary fermions (quantons) this results in a finite theory
without open parameters, which may be regarded as a fundamental description of
the strong interaction. 
Two potentials are deduced, a boson-exchange potential and one, which can be
identified with the known confinement potential in hadrons. This formalism has
been applied the mesonic systems $\omega(782)$, $\Phi(1020)$, $J/\psi(3097)$
and $\Upsilon(9460)$, for which a good description is obtained. 

The most important results are: 
1.~The confinement of hadrons is not due to colour, but is a general
property of relativistic bound states.
2.~Massive quarks in the Standard Model (QCD) are understood as effective
fermions with a mass given by the binding energy in the boson-exchange
potential. 

PACS/ keywords: 11.15.-q, 12.40.-y, 14.40.Cs/ Bound
state description of had-rons based on a second order Lagrangian with massless
fermions (quantons) and two-boson coupling. Confinement and boson-exchange
potential. Quarks understood as effective fermions with masses given 
by bound state energies. Mesonic systems $\omega(782)$, $\Phi(1020)$,
$J/\Psi(3097)$ and $\Upsilon(9460)$ studied. 
\end{abstract}

To study fundamental forces, nature provides us with hadrons and leptons,
which form the constituents of matter, but also with composite systems,
nuclei, atoms, and gravitational  states in form of solar and galactic
systems. For the description of these stable and massive systems as bound
states a relativistic theory is needed, since the elementary
constituents of these states are relativistic. However, relativistic
bound state problems are generally difficult to solve, see e.g.~Salpeter and
Bethe~\cite{Salbe}, and could not be tackled so far for particle
bound states (see the discussion in ref.~\cite{Glazek}).

Instead, powerful effective theories have been developed, which are contained
(except gravitation) in the Standard Model of particle physics~\cite{PDG}
(SM). In these divergent first order field theories particle bound states are
included effectively by a number of parameters including nine masses of
'elementary' particles. Only in QED bound states have been
calculated from the Coulomb potential. But also in this theory the
magnitude of the coupling constant $\alpha$ is not understood from first
principles. 

To understand the underlying mechanisms as well as the parameters needed in
first order theories, a more fundamental theory should exist, in which
these features are explained. If we demand further a real physical
understanding of the development of particle systems in 
the universe, this theory should most likely be finite, since nature develops
in a smooth way without singularities. Such a theory is expected to be based
on a Lagrangian including higher order fields. This can be seen for example
from the mass of light hadrons, which is much larger than the underlying quark
masses.

However, in spite of an evident need for higher order theories, there is a
strong belief that Lagrangians describing fundamental forces can be only first
order. This is correct for divergent field theories, since the inclusion of
higher order terms destroys the renormalisability of the theory. But this
argument is not valid, if a finite theory is constructed.  
Another argument against the use of higher order theories is that in such
theories Lagrangians with higher field derivatives~\cite{highd} are required,
which can lead to unphysical solutions (ghosts). But it should be 
realised that the problem is not the use of higher order theories in
principle, but to find a form of the Lagrangian, in which all important
criteria of relativistic theories are respected, like gauge invariance and
energy-momentum conservation. 

Recently, a second order theory has been developed by extending the QED
Lagrangian by boson-boson coupling~\cite{Mosneu}.  This formalism fulfills the
above criteria of a relativistic theory and can be regarded as a
fundamental description of the electric force in light atoms. However, for
hadronic bound states the requirements are still higher and ask also for
massless elementary fermions (quantons). Then, the mass of $q\bar q$ bound
states has to be entirely due to binding energy. 

An interesting question is, whether in a fundamental theory of
hadrons the colour degree of freedom is really needed (since hadrons are
colour neutral). The answer depends on confinement. If it would be due to
colour (as often assumed~\cite{Greensite}), a non-Abelian theory with colour
would be needed. However, already in the description of atomic systems a
confinement potential has been found~\cite{Mosneu}, suggesting that
confinement is a basic property of bound states of relativistic particles.   

The Lagrangian may be written in the form  
\begin{equation}
\label{eq:Lagra}
{\cal L}=\frac{1}{\tilde m^{2}} \bar \Psi\ i\gamma_{\mu}D^{\mu}
D_{\nu}D^{\nu} \Psi\ -\ \frac{1}{4} F_{\mu\nu}F^{\mu\nu}~,   
\end{equation}
where $\tilde m$ is a mass parameter and $\Psi$ in general a two-component
fermion field $\Psi=(\Psi^+$ $\Psi^o$) and $\bar 
\Psi= (\Psi^-$ $\bar \Psi^o$) with charged and neutral part. Vector boson
fields $A_\mu$ with charge coupling g are contained in the covariant derivatives
$D_{\mu}=\partial_{\mu}-i{g} A_{\mu}$ and the Abelian field 
strength tensor $F^{\mu\nu}=\partial^{\mu}A^{\nu}-\partial^{\nu}A^{\mu}$. 

We insert $D^{\mu}=\partial^{\mu}-i{g} A^{\mu}$ and
$D_{\nu}D^{\nu}=\partial_{\nu}\partial^{\nu}
-ig(A_{\nu}\partial^\nu+\partial_\nu A^\nu) -g^2 A_\nu A^\nu$ in
eq.~(\ref{eq:Lagra}) and obtain for the first term of ${\cal L}$
\begin{displaymath}
{\cal L}_1 = \frac{1}{\tilde m^2}\bar \Psi\ i\gamma_{\mu}D^{\mu}
D_{\nu}D^{\nu} \Psi = \frac{i}{\tilde m^2} \bar \Psi\ \gamma_{\mu}
\partial^{\mu} \partial_{\nu}\partial^{\nu} \Psi  
+\frac{g}{\tilde m^2} \bar \Psi\ \gamma_\mu A^{\mu}
\partial_{\nu}\partial^{\nu} \Psi
\end{displaymath}
\begin{displaymath}
+\frac{g}{\tilde m^2} \bar \Psi\ \gamma_\mu \partial^\mu
A_{\nu}\partial^{\nu} \Psi 
+\frac{g}{\tilde m^2} \bar \Psi\ \gamma_\mu \partial^\mu
\partial_{\nu}A^{\nu} \Psi 
-\frac{ig^2}{\tilde m^2} \bar \Psi\ \gamma_\mu A^\mu A_{\nu}\partial^{\nu} \Psi
\end{displaymath}
\begin{equation}
\label{eq:Lsum}
-\frac{ig^2}{\tilde m^2} \bar \Psi\ \gamma_\mu A^\mu \partial_{\nu}A^{\nu} \Psi
-\frac{ig^2}{\tilde m^2} \bar \Psi\ \gamma_\mu \partial^\mu A_{\nu}A^{\nu} \Psi
-\frac{g^3}{\tilde m^2} \bar \Psi\ \gamma_\mu A^\mu A_{\nu}A^{\nu} \Psi\ .
\end{equation}
The gauge condition $\partial_\mu A^\mu=0$ used for simpler Lagrangians (as in
QED) is replaced in our case by $\partial(\partial_\nu A^\nu)=0$. 

In eq.~(\ref{eq:Lsum}) the number of field derivatives and boson couplings
varies between the first and last term. This shows that the various terms are
related to different kinetic situations, pointing to a rather complex dynamics
of the system. 

Contributions to stationary solutions have been studied by using the standard
method of evaluating fermion matrix elements (or ground state expectation
values) of field operators~\cite{PS} derived from generalised Feynman diagrams. 
These can be written in the form 
${\cal M}(p'-p)=\ <g.s.|~K(q)~|g.s.>\sim \bar \psi(p')\ K(q)\ \psi(p)$, 
where $\psi(p)$ is a fermionic wave 
function $\psi(p)=\frac{1}{\tilde m^{3/2}} \Psi(p_1)\Psi(p_2)$ 
and $K(q=p'-p)$ a kernel, which is expressed by 
$K(q)=\frac{1}{\tilde m^{2(n+1)}}\ [O^n(q)\ O^n(q)]$, 
where n is the number of boson fields and derivatives in
eq.~(\ref{eq:Lsum}) (in the present case n=3).  

For the construction of stationary states we expect contributions
mainly from terms of the Lagrangian~(\ref{eq:Lsum}), which contain static
fields (without derivatives). This is the case only for the last term ${\cal
  L}_{1,8} = -\frac{1}{\tilde m^2} \bar 
\Psi\ g^3 \gamma_\mu A^\mu A_{\nu}A^{\nu} \Psi$ and leads to a matrix element
${\cal M}_{3g}$, which contains three boson fields on the right and left 
\begin{equation}
\label{eq:M3}
{\cal M}_{3g} = \frac{-{\alpha}^{3}}{\tilde m^8}~\bar
\psi(p')\gamma_\mu A^\mu(q)A_\nu(q)A^\nu(q)~
A_\sigma(q)A^\sigma(q)\gamma_\rho A^\rho(q) \psi(p) \ , 
\end{equation}
where $\alpha=g^2/4\pi$. A comparable matrix element in a first order theory
may be written in the form ${\cal M}_{f.o.} = \frac{-\alpha}{\tilde m^4}~\bar
\psi(p')\gamma_\mu A^\mu(q)~\gamma_\rho A^\rho(q) \psi(p)$,
giving rise to a (boson-exchange) interaction of vector structure $v_v(q) \sim
\alpha A_\mu(q) A^\rho(q)$ (but only for equal times of the two boson
fields, which means in the non-relativistic limit). Differently, in
eq.~(\ref{eq:M3}) three interactions of scalar and vector structure
$V_\mu^\nu(q) \sim \alpha A_{\mu}(q) A^{\nu}(q)$ are involved. 
In a dual picture the two boson fields, which appear twice (on the
left and right side of ${\cal M}_{3g}$) can be regarded (analoguous to the
fermion wave function $\psi(p)$) as bosonic (quasi) wave 
functions\footnote{leading to boson (quasi) densities $w^2(q)$ with dimension
  $[GeV]^2$.} $W_{\mu}^\nu(q)= \frac{1}{\tilde m} A_\mu(q) A^\nu(q)$. 
The fact that boson fields can be combined to wave functions leads quite
naturally to a finite theory, in which the wave functions are normalised. 

The physical picture of ${\cal M}_{3g}$ is that for the lowest energy state of a
relativistic bound state system the fermions interact only inside the
two-boson density (which leads to a boundary condition discussed below) and
feel therefore three interactions. 
A single boson-exchange interaction is possible only in dynamical situations,
see the terms 2-4 in eq.~(\ref{eq:Lsum}), which do not lead to a bound state
potential.   

The $\gamma$-matrices can be removed (contracted)
by adding a matrix element with interchanged $\mu$ and $\rho$ (according to
$\frac{1}{2} (\gamma_\mu\gamma_\rho+ \gamma_\rho\gamma_\mu) =g_{\mu\rho}$). 
Further, an equal time requirement of the two-boson fields (to reach overlap)
allows to replace all fermion four-vectors\footnote{in a ($t,\vec r$)
  representation} by three-vectors in momentum or r-space. Correspondingly,
the boson wave functions $W_\mu^\nu(q)$ and the remaining (boson-exchange)
interaction $V_\mu^\nu(q)$ are reduced to $w_{s,v}(q)$ and $v_{v}(q)\sim
w_{v}(q)$, which are two-dimensional. This yields   
\begin{equation}
\label{eq:M3f}
{\cal M}_{3g} = \frac{-{\alpha}^{3}}{\tilde m^5}~\bar
\psi(p')w_{s,v}(q)~v_{v}(q)~w_{s,v}(q) \psi(p) \ .
\end{equation}
Writing the matrix element by ${\cal M}_{3g} = \bar
\psi(p')\ V_{3g}(q)\ \psi(p)$ we obtain a three-boson potential
\begin{equation}
\label{eq:V3}
{ V}^{s,v}_{3g}(q)=\frac{-\alpha^3}{\tilde m^{2}}~w_{s,v}^2(q) v_v(q)\ .
\end{equation}
Fourier transformation to r-space leads to a folding
potential   
\begin{equation} 
\label{eq:vqq}
V^{s,v}_{3g}(r)= -\frac{\alpha^3 \hbar}{\tilde m} \int dr'
w_{s,v}^2(r')\ v_v(r-r')~.   
\end{equation}
Such a form has been used to describe elastic and inelastic hadron
processes~\cite{MoSDe}. 

The bosonic part of eq.~(\ref{eq:M3f}) can also be written in the form
of a matrix element, in which the wave functions $w_{s,v}(q)$ are
connected by $v_v(q)$
\begin{equation}
\label{eq:P2g'}
{\cal M}^{g} =\frac{-\alpha^{3}}{\tilde m^{2}}
\ w_{s,v}(q) v_{v}(q) w_{s,v}(q) .
\end{equation}
This matrix element shows binding of two bosons in the potential $v_{v}(q)$,
consequently $\partial^2 w_s(q)$ is related to their
kinetic energy. The contribution from the vector part  $\partial^2 w_v(q)$
cancels out as a consequence of the gauge condition. Below it will be
shown that this implies also the existence of a static two-boson potential
$V_{2g}(q)$.  

From the general structure of the fermion matrix element in eq.~(\ref{eq:M3f}) 
one can see that there are two fundamental s-states (with quantum
numbers $(J^\pi=1^-)$) with scalar and vector boson wave functions
$w_{s,v}(r)$ and corresponding fermion wave functions\footnote{for the radial
  wave functions $\bar \psi(r)=\psi(r)$.} $\psi_{s,v}(r) \sim w_{s,v}(r)$
normalised by $4\pi\int r^2dr~\psi_{s,v}(r)=1$. Further, there are 
0$^+$ states with p-wave functions $\psi_{L=1}(r)$, which are not considered in
the present paper.  

Orthogonality of the total wave functions requires that the boson
wave functions $w_{s}(r)$ and $w_{v}(r)$ are orthogonal and lead to a
vanishing radial matrix element  
\begin{equation}
\label{eq:ortho}
<r^2_{w_s,w_v}>=\int r^3dr~w_s(r) w_v(r)=0 \ .
\end{equation}
To satisfy this condition, $w_{v}(r)$ may be written in the form of a
p-wave function 
\begin{equation}
\label{eq:spur}
w_{v}(r) = w_{v,o}~[w_s(r)+\beta R\ \frac{d w_s(r)}{dr}]~,
\end{equation}
where $w_{v,o}$ is obtained from the normalisation $2\pi \int r dr\ w_v^2(r)
=1$ and $\beta R$ from eq.~(\ref{eq:ortho}). 

To evaluate the potentials $V^{s,v}_{3g}(r)$, the boson wave functions
$w_{s,v}(r)$ have to be determined. To achieve this, a boundary condition can
be formulated by requiring that the interaction takes place inside the volume
of the strongest bound state. As a consequence, the corresponding
boson-exchange potential~(\ref{eq:vqq}) should be proportional to the density
$\psi^2(r)$ $\sim w^2_s(r)$, leading to 
\begin{equation}
\label{eq:conr}
{c}\ w^2_s(r) \sim |V^v_{3g}(r)| \ .  
\end{equation}
Both constraints~(\ref{eq:ortho}) and (\ref{eq:conr}) can be satisfied by 
a boson wave function of the form 
\begin{equation}
\label{eq:wf}
w_s(r)=w_{s_o}\ exp\{-(r/b)^{\kappa}\} \ , 
\end{equation} 
where $w_{s_o}$ is the normalisation factor $1/(4\pi \int r
dr\ w_s^2(r))$. 
The boson-exchange interaction $v_v(r)$ is given by $v_v(r)=\hbar~w_v(r)$.
\vspace{0.5cm}

To generate a stable bound state, the potential $V^{s,v}_{3g}(r)$ is not
sufficient to keep the bosons confined. The other terms of
the Lagrangian~(\ref{eq:Lsum}) show kinematic situations, in which
bosons and/or fermions are in motion. Nevertheless, term 6 may be
written in the form ${\cal L}_{1,6} = 
-\frac{ig^2}{\tilde m^2} \bar \Psi\ \gamma_\mu A^\mu (\partial_{\nu}A^{\nu})
\Psi -\frac{ig^2} {\tilde m^2} \bar \Psi\ \gamma_\mu A^\mu
A_{\nu}~\partial^{\nu} \Psi$ and gives rise to another bound state potential.  

The first term of ${\cal L}_{1,6}$ leads to 
\begin{equation}
\label{eq:M2}
{\cal M}_{2g} =\frac{\alpha^{2}}{\tilde m^8} \bar \psi(p')\gamma_\mu
A^{\mu}(q) (\partial_\nu A^{\nu}(q))~\gamma_\rho A^{\rho}(q)
(\partial_\sigma A^{\sigma}(q)) \psi(p)\ .
\end{equation}
Using the gauge condition we can write $(\partial_\nu
A^{\nu}(q)) (\partial_\sigma A^{\sigma}(q))=\frac{1}{2} \partial_\nu
[\partial_\sigma (A_\mu A^\mu) ^\sigma]^\nu$.
After contracting the $\gamma$-matrices and reducing the fermion and boson
vectors by one dimension as discussed for ${\cal M}_{3g}$, this yields 
\begin{equation}
\label{eq:P2g}
{\cal M}_{2g} =\frac{\alpha^{2}}{2\tilde m^{6}} \bar \psi(p')
\ w_s(q) \partial^2 w_s(q)\ \psi(p)\ .
\end{equation}
Since the two bosons are bound, see eq.~(\ref{eq:P2g'}), $\partial^2 w_s(q)/
2\tilde m$ is related to their kinetic energy distribution. According to the
virial theorem this implies also the existence of a static two-boson potential
$V_{2g}(q)$.  

In a transformation to r-space the bosonic part of eq.~(\ref{eq:P2g}) gives
rise to a Hamiltonian of the form
\begin{equation}
\label{eq:H}
-\frac{\alpha^2 \tilde m <r^2_{w_s}> F_{2g}}{4}~\Big (\frac{d^2 w_s(r)}{dr^2} + 
  \frac{2}{r}\frac{d w_s(r)}{dr}\Big ) +V_{2g}(r)~w_s(r) =
E_i~w_s(r)~,
\end{equation} 
where the factor $F_{2g}$ is due to the Fourier transformation of the boson
kinetic energy, $<r^2_{w_s}>$ the radius square of the boson density and
$w_s(r)$ the Fourier transform of $w_s(q)$. The potential $V_{2g}(r)$ is 
given by 
\begin{equation}
V_{2g}(r)= \frac{\alpha^2\tilde m <r^2_{w_s}> F_{2g}}{4}\ \Big
(\frac{d^2 w_s(r)}{dr^2} + 
  \frac{2}{r}\frac{d w_s(r)}{dr}\Big )\frac{1}{\ w_s(r)}+E_o\ , 
\label{eq:vb}
\end{equation}
where $E_o=0$ is used to make a connection to the vacuum (state
without binding between the quantons and therefore $E_{vac}=0$). 
A similar potential involving $w_v(q)$ deduced from ${\cal L}_{1,7}$ yields
negligible contribution to the binding energy. All other terms of the
Lagrangian~(\ref{eq:Lsum}) do not contribute to bound state potentials. 

The implications of using massless fermions are very strong and can be
summarized as follows: 
First, the vacuum of the theory is the absolute vacuum with average energy 
$E_{vac}=0$. This is consistent with the low energy density
of the universe deduced from astrophysical observations. Second,  
the lowest energy solution in $V_{2g}(r)$  is the vacuum and
therefore $E_o=E_{vac}=0$. By this condition the absolute height of $V_{2g}(r)$ is
fixed.  Third, by rewriting eq.~(\ref{eq:P2g}) in the form ${\cal M}_{2g} = 
\frac{\alpha^{2}}{2\tilde m^6} w_s(q) \{ \bar \psi(p') \psi(p)\} \partial^2
w_s(q)$, one can see that fermion-antifermion pairs can be created during 
the dynamical overlap of two fluctuating boson fields. By this mechanism stable
particles can be created out of the absolute vacuum. These facts are
consistent with the requirement for a fundamental theory.

An important fact is that $V_{2g}(r)$ can be identified with the confinement
potential in hadron potential models~\cite{qq}. This will be
shown in a comparison with the confinement potential from lattice QCD
simulations~\cite{Bali} and the discussion of quark masses. 

$V_{2g}(r)$ can also be written in a different form 
\begin{equation}
V_{2g}(r)= \frac{\alpha^2 \hbar^2 F_{2g}}{4\tilde m}\ \Big
(\frac{d^2 w_s(r)}{dr^2} + 
  \frac{2}{r}\frac{d w_s(r)}{dr}\Big )\frac{1}{\ w_s(r)}+E_o\ . 
\label{eq:vb1}
\end{equation}
This leads to the condition 
\begin{equation}
Rat=\frac{\hbar^2}{\tilde m^2 <r^2_{w_s}>} =1 \ .
\label{eq:ravb}
\end{equation}
A last constraint is related to energy-momentum conservation
in relativistic systems, indicating that for binding in
$V_{3g}^s(r)$ the total energy of the system is not increased, the negative
fermion and boson binding energies $E_f^s$ and $E_g$ have to be compensated by
the root mean square momenta of the corresponding potentials   
\begin{equation}
<q^2_{V_{3g}}>^{1/2}+<q^2_{v_v}>^{1/2} = -(E_f^s+E_g) \ . 
\label{eq:massq}
\end{equation}
However, for the confinement potential $V_{2g}(r)$ this condition is not valid.
Therefore, the constraint~(\ref{eq:massq}) can be applied only for the
binding potentials $V(q)=V_{3g}(q)$ and $v_v(q)$ with $<q^2_{V}>= f_{red} \int
  dq~q^3 V(q)/ \int dq~q V(q)$ and $f_{red}=E_f^s/(E_f^s+E_g)$.
\vspace{0.3cm}

The fermion mass of the system is defined by the energy to balance binding  
\begin{equation}
\label{eq:mass}
M^{s,v}_{n}=-E_{f_{s,v}}^{3g}+E_{f_n}^{2g} \ , 
\end{equation}
where $E_{f_{s,v}}^{3g}$ is the negative binding energy in $V^{s,v}_{3g}(r)$ 
(for these potentials only the lowest state is discussed
here) and $E_{f_n}^{2g}$ are positive binding energies for different (excited)
states in $V_{2g}(r)$. This shows two types of mass generation, binding in the
Coulomb like potential $V_{3g}(r)$ and dynamical mass generation in $V_{2g}(r)$. 

In the whole formalism there are finally four constraints~(\ref{eq:ortho}),
(\ref{eq:conr}), (\ref{eq:ravb}) and (\ref{eq:massq}), by which {\bf all}
open parameters, shape parameter $\kappa$, slope (or size) parameter $b$ 
and the coupling constant $\alpha$, are determined within rather small
ambiguities. In addition, the different flavour states in the quark
model\footnote{the notion of flavour from the quark model is kept in the
  present approach to characterise systems of different slope parameter $b$.}
can be related by 
a vacuum sum rule similar to that applied in ref.~\cite{Mosneu}, which
indicates that in principle a complete solution of the relativistic bound
state problem for all states is achieved. 
Below it will be shown that the need for massless elementary fermions in
the present formalism is entirely consistent with the requirement of 
finite quark masses in the SM.

-----------

The above formalism has been applied to $q\bar q$ mesons (of different flavour
structure in the quark model) $\omega(782)$, $\Phi(1020)$, charmonium
$J/\Psi(3097)$ and bottonium $\Upsilon(9460)$ including excited states.  
The potentials $V_{3g}(r)$ and $V_{2g}(r)$ have been determined by adjusting
the open parameters to fulfill the constraints discussed above. Remaining 
uncertainties have been reduced by fine-adjustment of the factor $F_{2g}$ in
the confinement potential $V_{2g}(r)$ to fit the spectrum of radial
excitations.

Results on the radial dependence of densities and potentials are
given in fig.~1 for the $\omega(782)$ system. In the upper
part the interaction $v_v(r)$ is given by the solid line. Compared to the
Coulomb potential $v_{coul}(r)=\hbar/r$ (dot-dashed line) there are no
divergencies for $r\to 0$ and $\infty$, in agreement with the demand of a
finite theory. 

In the middle part a comparison of the density $w_s^2(r)$ (dot-dashed line)
with the potentials $V^s_{3g}(r)$ (dashed line) and $V^v_{3g}(r)$ (solid line)
is made. We see that condition~(\ref{eq:conr}) for the vector potential is
reasonably well fulfilled at larger radii. This indicates that the bosonic
wave function 
$w_s(r)$ is well described by the radial form in eq.~(\ref{eq:wf}) and that
also relation~(\ref{eq:spur}) between $w_s(r)$ and $w_v(r)$ is correct. 
 
\begin{table}
\caption{Results for mesonic systems, $\omega(782)$, $\Phi(1020)$,
  $J/\psi(3097)$, and $\Upsilon(9460)$ including excited states, in comparison
  with the data~\cite{PDG}. Masses and binding energies are given in GeV, $b$
  in fm, and the mean radius squares in fm$^2$. $\alpha_{eq}$ is the
  equivalent coupling constant in the Coulomb potential.}   
\begin{center}
\begin{tabular}{c||cccc|c|ccc}
System &$M^s_1$& $M^s_{2}$ &$M^s_{3}$&$M^s_{4}$ & $M^v_1$& $ M_1^{exp}$& $
M_2^{exp}$ & $M_3^{exp}$   \\   
\hline
$\omega$ & 0.78 & 1.42  & 1.93 &      & 1.3 & 0.782 & 1.42$\pm$0.03  \\  
$\Phi$  & 1.02 & 1.68  & 2.20 &      & 2.0 & 1.02  & 1.68$\pm$0.02  \\
$J/\psi$      & 3.10 & 3.69  & 4.16 & 4.58 & 8.1 & 3.097 & 3.686 &
4.16$\pm$0.02   \\  
$\Upsilon$ & 9.46 & 10.02 & 10.46 & 10.8 & 26.6 & 9.46 & 10.023 & 10.355 \\
\end{tabular}
\begin{tabular}{c||ccc|cccc|c}
System & $\kappa$ & $b$ & $\alpha$ & $\alpha_{eq}^*$ & $E_{g}$
& $E^s_{3g}$ & $m_{quark}$ &
$<r^2_{w_s}>$ \\ 
\hline
$\omega$ & 1.4 & 0.589 & 0.65 & 0.05 & -0.026 & -0.013 & 0.0065 & 0.256 \\ 
$\Phi$   & 1.4 & 0.450 & 1.50 & 0.65 & -0.43 &  -0.216 & 0.108  & 0.150 \\
$J/\psi$ & 1.4 & 0.148 & 2.32 & 2.46 & -4.82 & -2.42 & 1.21 & 0.016  \\ 
$\Upsilon$ & 1.4& 0.049 & 2.49 & 3.04 & -18.4 & -9.0 & 4.5 & 0.0017 \\ 
\end{tabular}
\end{center}
~\hspace{1.1cm} * $\alpha_{eq}=\sum_{s,v} \int dr\ V_{3g}^{s,v}(r)/\int
dr\ V_{coul}(r)$ 
\end{table}
In the lower part of fig.~1 the deduced confinement potential $V_{2g}(r)$ is
shown. It is characterized by a close to linear form at larger radii, as
expected from ref.~\cite{qq,Bali}. Resulting masses and parameters for 
different systems are given in table~1. Although the binding energies are
quite different, in all cases a satisfactory agreement of the various
quantities is obtained, which fulfill all boundary conditions. A similar plot
as in fig.~1 is shown for the bottonium system in
fig.~2. Apart from a very different radial extent of the two systems the only
important difference is the relative size of the confinement potential, which
is drastically reduced for the heavy system due to a very different dynamics.  

A comparison of the deduced confinement potential $V_{2g}(r)$ with the lattice
QCD simulations of Bali et al.~\cite{Bali} (solid points with error bars) is 
shown in fig.~2. This potential has the same form as the confinement potential
$V_{conf}(r)\sim -\alpha/r+l\cdot r$ deduced from potential
models~\cite{qq}. The fact that very similar results are deduced from theories
with and without colour indicates clearly that the confinement of hadrons is
not due to colour (as assumed in ref.~\cite{Greensite} but without clear
understanding of the mechanisms involved), but represents a
general property of relativistic bound states. 

The question of a vector or scalar structure of
the confinement potential can be studied by looking at the splitting of p-wave
states in charmonium and bottonium, see ref.~\cite{question}. From 
the existing data neither a vector nor a scalar structure is found, 
supporting strongly a derivative structure of the potential $V_{2g}(r)$, as
found in the present approach.
 
An important point is the need for finite quark masses in the SM (QCD), which
should be understood in the present more fundamental approach. 
These masses have been estimated in different models, as e.g.~in QCD inspired
potential models~\cite{qq} (more details can be found in ref.~\cite{PDG}). 
The empirical form $V_{conf}(r)\sim -\alpha/r+ l\cdot r$ assumed in
these models is consistent with $V_{2g}(r)$; therefore, the quark masses
have to be related to the binding energy in $V^s_{3g}(r)$. This leads to the
relation\footnote{this expression is independent of using an Abelian or
  non-Abelian structure of the Lagrangian. In an Abelian theory massive
  ``quarks'' have the same charge as the quantons in eq.~(\ref{eq:Lagra}).}    
\begin{equation}
\label{eq:mquark}
m_{quark}=-\frac{1}{2}\ E_{3g}^s  \ .
\end{equation}
The resulting quark masses are given in table~1 and are compared with the
extracted masses~\cite{PDG} in fig.~3. An excellent agreement is
obtained. This is clear indication that the need for massive fermions (quarks)
in the first order theory (QCD) is perfectly consistent with the assumption of
massless elementary fermions in the present approach. Thus, the quarks
can be understood as {\bf effective fermions} with masses related to the
binding energy in the bound state potential $V_{3g}^s(r)$.  

Concerning an interpretation of the quark masses as due to the
Higgs-mechanism~\cite{Higgs}, such an explanation (which demands an extra high
energetic background field) is not needed. Further, the flavour structure of
mesons comes out naturally in the present approach. Therefore, supersymmetric
extensions of the SM, which predict a new regime of super-symmetric
particles at high energies, are also not needed. This confirms the general
view that a fundamental theory must have a very simple structure.
\vspace{0.5cm}
\newpage

In conclusion, although the SM yields an excellent description of many
particle properties, it is an effective theory with parameters (quark masses),
which are not well understood from first principles. To get a correct
understanding of the nature of these parameters, a more fundamental theory of
hadrons is required. This has been achieved in the present approach, in which
the colour degree of freedom as well as Higgs and supersymmetric fields are
not needed. Preliminary results from an application to different fundamental
forces can be found in ref.~\cite{Moneu5}. 
\vspace{0.9cm}

The author is very grateful to many colleagues for fruitful discussions,
valuable comments and the help in formal 
derivations. Special thanks to P.~Decowski and P.~Zupranski for numerous
conversations and encouragements and B.~Loiseau for his help with the
formulation of the Lagrangian and matrix elements.

\newpage
\begin{figure} [ht]
\centering
\includegraphics [height=16.7cm,angle=0] {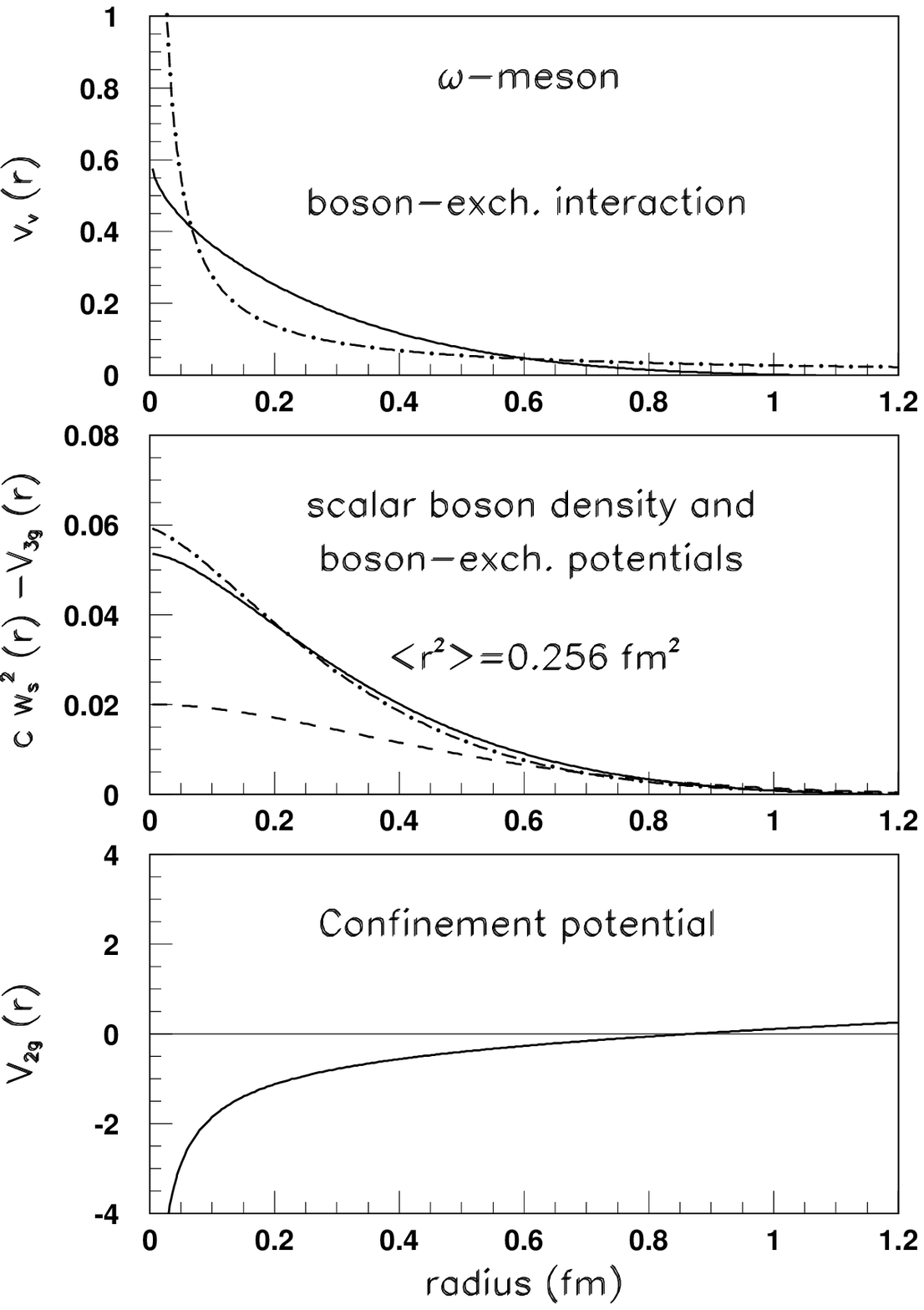}
\caption{Self-consistent solution for the $\omega(782)$ meson system.
  \underline{Upper part:} Interaction $w_v(r)$ in comparison with
  the Coulomb potential, given by solid and dot-dashed
  lines, respectively.
 \underline{Middle part:} Bosonic density $w_s^2(r)$ and potential
 $|V^{v}_{3g}(r)|$ given by the overlapping dot-dashed and
 solid lines, respectively, matched by the condition~(\ref{eq:conr});
 $|V^{s}_{3g}(r)|$ is shown by dashed line.
 \underline{Lower part:} Deduced confinement potentials $V_{2g}(r)$. }  
\label{fig:g1ex2g}
\end{figure}

\begin{figure} [ht]
\centering
\includegraphics [height=16.7cm,angle=0] {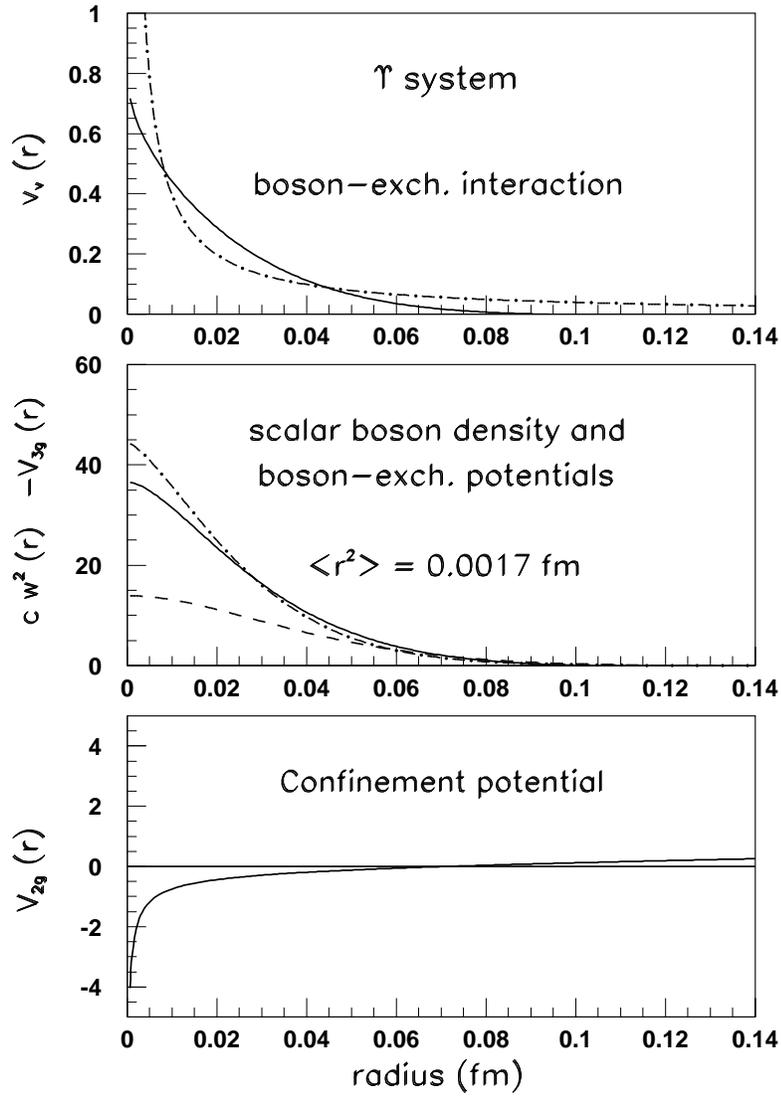}
\caption{Same as fig.~1 for the bottonium system
  $\Upsilon(9460)$. } 
\label{fig:g1ex2c}
\end{figure}

\begin{figure} [ht]
\centering
\includegraphics [height=15cm,angle=0] {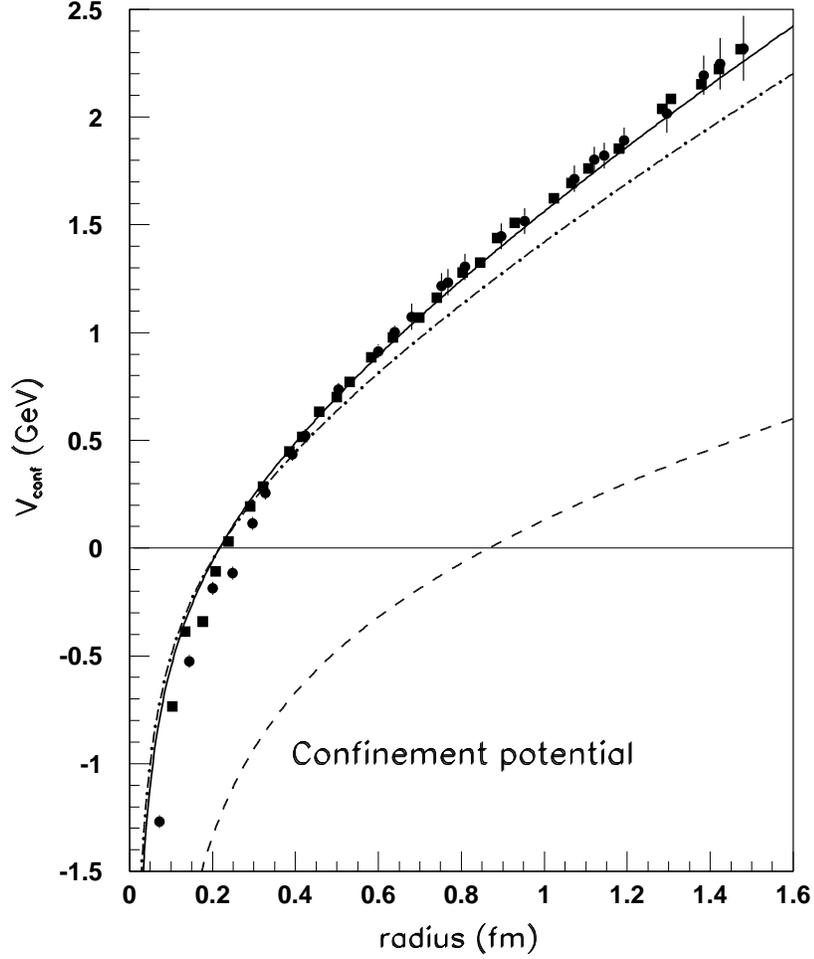}
\caption{Comparison of the confinement potential with lattice QCD
  calculations. $V_{2g}(r)$ calculated for the two mesonic systems
  $\omega(782)$ and charmonium $J/\psi$, given by dashed and dot-dashed line,
  respectively. The latter, multiplied with a factor 1.1 (solid line) shows an
  excellent agreement with lattice gauge
  simulations~\cite{Bali} (solid points). }
\label{fig:confine}
\end{figure} 

\begin{figure} [ht]
\centering
\includegraphics [height=12.0cm,angle=0] {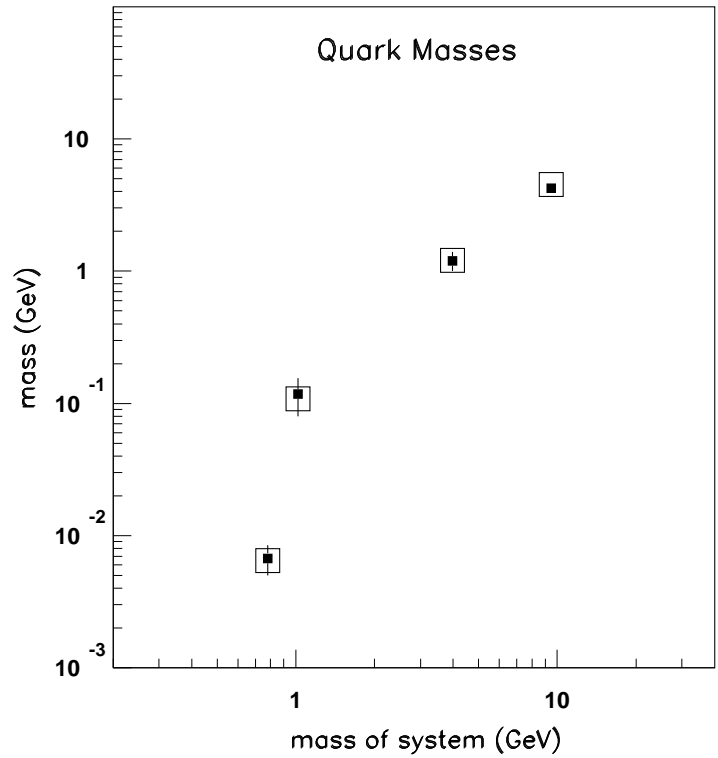}
\caption{Quark masses as a function of the g.s.~masses. The open squares show
  the present results using eq.~(\ref{eq:mquark}), the solid squares with
  error bars give the extracted values from other sources~\cite{PDG}. }
\label{fig:quarkm}
\end{figure} 

\end{document}